\documentclass[preprint,showpacs,preprintnumbers,amssymb]{revtex4}

\usepackage{graphicx} 
\begin{document}
\title{Localized Electron States Near a Metal-Semiconductor Nanocontact}

\author{Denis O. Demchenko and Lin-Wang Wang}
\affiliation{Lawrence Berkeley National Laboratory, Berkeley, California 94720}
\date{\today}

\begin{abstract}
The electronic structure of nanowires in contact with metallic electrodes of experimentally relevant sizes is calculated by incorporating the electrostatic polarization potential into the atomistic single particle Schr\"odinger equation. We show that the presence of an electrode produces localized electron/hole states near the electrode, a phenomenon only exhibited in nanostructures and overlooked in the past. This phenomenon will have profound implications on electron transport in such nanosystems. We calculate several electrode/nanowire geometries, with varying contact depths and nanowire radii. We demonstrate the change in the band gap of up to 0.5 eV in 3 nm diameter CdSe nanowires and calculate the magnitude of the applied electric field necessary to overcome the localization. 
\end{abstract}
\maketitle
Furture nanoelectronics will depend on electron/hole transport in a nanostructure, and crossing the nanostructure/metal electrode interface. 
Understanding the properties of such nanocontacts, especially new phenomena unique to the nanocontacts is thus of paramount importance to the future of nanoelectronics and it is currently an intensively studied subject. 
Significant progress has been made in recent years in fabricating variety of nanostructures and nanocontacts ranging from 0D structures like isolated quantum dots, to more sophisticated structures, such as tetrapods, nanoribbons, nanorods and nanowires, etc. \cite{tetrapods,branched,nanoribbon}. 
Transport measurements in such nanostructures are often conducted by contacting  a semiconductor nanostructure with large metallic electrodes, 
often deposited on top of the nanostructure.\cite{Alivisatos_dot_transistor,Salmeron,dumbells,ZnO_rod,Alivisatos_tetrapod_transistor,nanowire_transistor,Zhang_nanotube,diode} 
Hitherto, theoretical interpretation of such experiments has often been based on transport and electronic structure calculations with isolated nanostructures and ignoring influence of the electrodes. Even if the electrode is included in the calculation, its effects are often described by a short range correlation theory, such as local density approximation (LDA). 
For example, a nanocontact between an extremely small Si nanowire (with a diameter of 5 \AA) and Al electrode has been calculated by U. Landman et al. \cite{Si_nanowires}. However, due to the use of LDA, the authors did not find the localization effects shown in this letter. 
The existence of the electrode will introduce the long range correlation effects well into the semiconductor nanowire, causing the change of its electronic structure. The long range correlation effects cannot be described by LDA, they can only be addressed by many-body theories, like the $GW$ method, albeit the high cost of the $GW$ method prevents it from being used to calculate the systems we discuss here. Fortunately, in the static approximation, the long range correlation effect allows a simple classical interpretation. \cite{LW_electrod} It can be represented by an additional potential in the single particle Schr\"odinger equation, corresponding to a classical surface polarization potential (or image potential for metal). 
As we show below, this surface polarization potential is surprisingly large, particularly for small nanostructures, where it alters the familiar electronic structure and gives rise to a nanocontact phenomenon: electrode induced wavefunction localization. This localization disappears when the system becomes large, which explains why it has been overlooked in the past.

In this letter we study the common case of the CdSe nanowire attached to a generic metallic contact. In practical calculations we use a long nanorod to represent a nanowire. 
The physical properties of CdSe nanorods such as optical spectroscopy, conductivity, electric dipole, etc., have been extensively studied in the past decade \cite{CdSe_rod1,CdSe_rod2,CdSe_rod3,CdSe_rod4} both experimentally and theoretically. 
The charging properties of isolated nanostructures (which are dependent on the surface polarization potential) have been studied theoretically, for quantum dots \cite{charging_dot1,charging_dot2,Coulomb_blokade,dot_effective_mass}, and tetrapods \cite{LW_electrod}. 
It has been demonstrated that 
the surface polarization potential (when the set up does not include a metallic electrode) plays an important part in the quantum dot charging energy.\cite{LW_electrod} Here we study the electronic properties of the CdSe nanorods in contact with metallic electrodes taking into account surface polarization potential. We use semi-empirical pseudopotential method (SEPM) to describe electron Schr\"odinger equation \cite{SEPM}. We assume that the metal electrode makes a Schottky contact with semiconductor nanorod as suggested in Ref. \onlinecite{F.Leonard_Schottky} (to avoid possible complications due to charge transfer). 

Within the SEPM we ordinarily solve the single-particle Schr\"odinger equation using planewave basis sets
\begin{equation}
 \{ - {1\over 2} \nabla^2 + V \} \psi_i(r) = E_i \psi_i(r)
\label{eq: Schodinger}
\end{equation}
where, $V$ is a potential which includes both local (sum of the screened atomic pseudopotentials) and nonlocal ($s,p,d$, and spin-orbit coupling) parts. The pseudopotentials are fitted to match the bulk experimental band structure.  
The potential $V$ is bulk-like inside and zero outside the nanorod. The dangling bonds at the nanorod surface are passivated with a model ligand potentials to eliminate the midgap surface states. Once the potential $V$ for the entire nanorod is constructed, the Eq. \ref{eq: Schodinger} can be used to solve for the eigenenergies and the wavefunctions in the nanorod. In practice, we use the folded-spectrum method (FSM) \cite{folded_spectrum} in order to solve for only a few states in the valence and conduction band adjacent to the bandgap. 
The use of FSM allows us to perform calculations of large nanostructures containing tens of thousands of atoms, and approach experimentally relevant sizes range. 

\begin{figure}[t]
\includegraphics[width=3.2in]{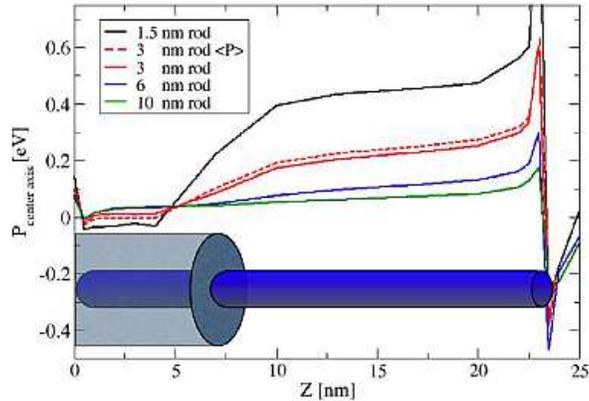}
\caption{(color online) Polarization potential $P(z)$ as a function of the $z$-coordinate running along the nanorod center axis (in case of the 3 nm rod, also the weighted average $\langle P(z) \rangle $). The nanorod is embedded into the electrode by a quarter of its length, nanorods of diameter 1.5, 3, 6, and 10 nm are used. The inset shows an example of the modeled electrode and CdSe nanorod used in the present work (electrode size is not to scale, which has a diameter of 23 nm). The dielectric constant is $\epsilon=10$ in the CdSe nanorod and $\epsilon \rightarrow \infty$ in the metal electrode. 
\label{fig: Polarization}}
\end{figure}

In the presence of the electrode, the surface polarization potential due to interaction of an electron with its image charge becomes significant. Proper treatment of these long range correlations requires a many-body approach, such as $GW$ method. However, it has been shown \cite{LW_electrod} that (under a static approximation) the self-energy potential in $GW$ equation can be split into a bulk potential and a surface polarization potential. The $GW$ equation is then split into two parts, the bulk Hamiltonian (analogous to Eq.\ref{eq: Schodinger}) and the surface polarization part, which is reduced to an electrostatic surface polarization potential $P(r)$. Thus, the $GW$ equation can be approximated as
\begin{equation}
 \{ - {1\over 2} \nabla^2 + V \pm P(r) \} \psi_i(r) = \epsilon_i \psi_i(r)
\label{eq: quasiparticle}
\end{equation}
where + and - are applied to electron and hole states, respectively, and $\epsilon_i$ are the quasiparticle energies. 
In the nanostructure the polarization potential $P(r)$ can be derived from the many-body $GW$ model, \cite{LW_electrod} which results in the same form as in the early electrostatic model of Brus \cite{Brus1,Brus2}
\begin{equation}
P(r) = {1\over 2} \lim_{r^\prime \rightarrow r} [W_{nano}(r^\prime,r)-W_{bulk}(r^\prime,r)]
\label{eq: polarization}
\end{equation}
where $W_{nano}(r^\prime,r)$ is the screened electrostatic potential in the nanorod at $r^\prime$ created by a test charge at $r$, and $W_{bulk}(r^\prime,r)$ is such potential in the bulk. 
In order to find the electrostatic potential $W(r^\prime,r)$ we solve the Poisson equation for the structure shown in the inset to Fig.\ref{fig: Polarization} (electrode size not to scale) 
\begin{equation}
\nabla [\epsilon(r)\nabla\phi(r)]=4\pi\rho(r)
\label{eq: Poisson}
\end{equation}.
The dielectric function $\epsilon(r)$ is modeled following our study of dielectric response of a nanostructure \cite{LW_dielectric_dot}, which equals $\epsilon_{CdSe}=10$ at the center of the nanorod, and approaches 1 near the surface of the rod. 
We use 
Dirichlet boundary conditions of $1/r$ at the domain boundary and
and obtain $P(r)$ which is then added to the previously computed CdSe nanorod potential $V(r)$ in the Eq. \ref{eq: quasiparticle}, for the subsequent SEPM calculation of the electronic structure. 

We calculated surface polarization potential $P(r)$ for nanorods length of 23 nm, and the diameters of 1.5, 3, 6, and 10 nm. The diameter of the electrode was much larger than that of the nanorod ($d_{electr}$=23 nm) in order to minimize the influence of a specific electrode geometry and mimic the situation of many experimental nanocontact setups. The SEPM calculations were performed for CdSe nanorods in the wurtzite crystal structure, length of 23 nm, and 3 nm diameter, with total number of atoms of 5434. The polarization potential $P(r)$ was generated for a) the nanorods embedded into a metal electrode by one half and one quarter of its length, b) nanorod and the electrode in contact, c) nanorod and the electrode separated by a 1 nm layer of vacuum, and d) a free standing nanorod. 

Figure \ref{fig: Polarization} shows the polarization function $P(z)$ as a function of the $z$-coordinate running along the nanorod central axis, for nanorod diameter ranging from 1.5 to 10 nm, in the case of the nanorod embedded into the electrode by a quarter of its length. For the 3 nm rod we also computed the weighted average $\langle P(z) \rangle $ as 
\begin{equation}
\langle P(z) \rangle = \int \vert \psi_i(x,y) \vert^2 P(x,y,z) dx dy
\label{eq: weighted}
\end{equation}
where $\vert \psi_i(x,y) \vert^2=\int \vert \psi_i(x,y,z) \vert^2 dz$, and $\psi_i(x,y,z)$ is  taken either as a conduction band minimum (CBM) or valence band maximum (VBM) wavefunction (they do not make any practical difference). Thus, $\langle P(z) \rangle $ is a measure of the effective influence of the $P(r)$ on the relevant wavefunctions. For the 3 nm rod the comparison of the central axis and weighted average polarization functions shows that the central axis $P(z)$ is a good measure of the weighted average $\langle P(r) \rangle$. 

The large $P(r)$ introduced by the presence of an electrode is evident quarter length along the rod. The electrode influence decreases with increased nanorod diameter. At 10 nm diameter a drop of $P(r)$ near the electrode is almost indiscernible. This indicates that the localization effect which will be discussed below does not exist in a macroscopic bulk contact, which is why this has been overlooked before. 
However, for small nanorods the effect is surprisingly large, the $P(r)$ provides a strong confining potential for electrons or holes (about 0.5 eV for 1.5 nm nanorod), and leads to an electrode induced electron/hole localization. 

\begin{figure}[t]
\includegraphics[width=3.2in]{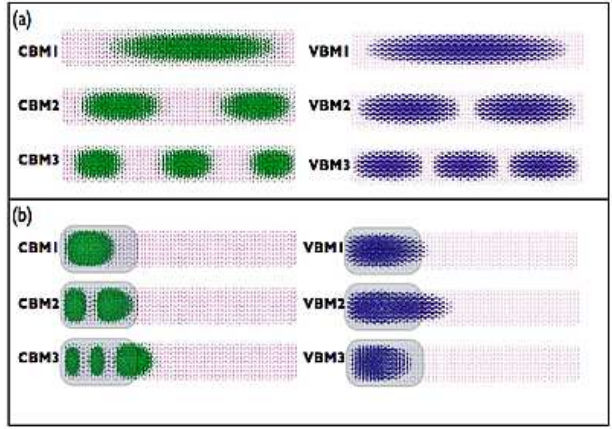}
\caption{(color online) Contour plot of the lowest three CBM wavefunctions of the 3 nm diameter CdSe nanorod, (a) CBM wavefunctions in the free standing nanorod, (b) CBM wavefunctions in the nanorod with an electrode covering one quarter of the rod on the right side. The semitransparent rectangles indicate the position of the electrode. 
\label{fig: wavefunctions}}
\end{figure}

Figure \ref{fig: wavefunctions} shows the real space contour plots of the three wavefunctions adjacent to the bandgap in the 3 nm CdSe quantum rod conduction and valence band. 
In Fig. \ref{fig: wavefunctions}(a) three wavefunctions are calculated for a free standing CdSe nanorod, 
while for Fig. \ref{fig: wavefunctions}(b) nanorod is embedded into the electrode (indicated by the semi-transparent rectangles) by one quarter of its length. In the latter case both CBM and VBM  wavefunctions are localized by the polarization potential induced by the electrode. 
While for CBM states, the wavefunctions are just shrunk in the $z$-direction, for VBM states the second and third states show different nodal structure in comparison with their free rod counterparts, indicating strong state mixing and crossing. 
Due to the electron and hole state localization, the transport properties of small nanorods measured with attached electrodes will be different from familiar free electron-like picture, and more resemble the case of the Coulomb blockade.

\begin{figure}[t]
\includegraphics[width=3.2in]{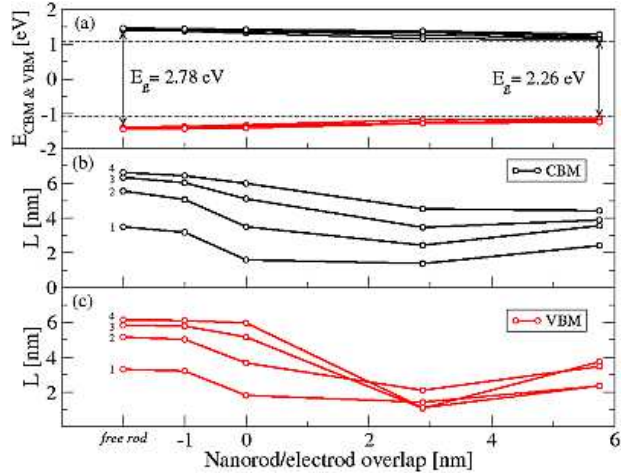}
\caption{(color online) Evolution of the electronic states for the 3nm diameter CdSe nanorod  as a function of the nanorod/electrode overlap, (a) CBM and VBM eigenvalues (bandgap), dashed lines indicate the bandgap computed ignoring surface polarization potential; (b) and (c) localization function (defined in the text) for the CBM and VBM states respectively. 
\label{fig: Gap_localiz}}
\end{figure}

Another important consequence of the electrode induced polarization potential is the change in the value of the bandgap. Figure \ref{fig: Gap_localiz} shows the VBM and CBM states evolution as a function of the overlap between the nanorod and the electrode for the 3 nm diameter CdSe nanorod. In this case the value of the bandgap is reduced by approximately 0.5 eV (Fig.\ref{fig: Gap_localiz}(a)), from 2.78 eV to 2.26 eV as the rod is embedded into the electrode. The bangap value changes slightly when the nanorod and the electrode are separated by the 1 nm layer of vacuum ($E_g=2.74$ eV). Once the electrode is in contact with the nanorod the change is more pronounced ($E_g=2.64$ eV) and becomes more significant with increasing the electrode/nanorod overlap, eventually saturating. 
In the absence of the electrode, the polarization potential $P(r)$ of a free standing nanorod induces the bandgap increase of 0.54 eV (from 2.24 eV to 2.78 eV) in comparison with the gap calculated ignoring the surface polarization (dashed lines in Fig.\ref{fig: Gap_localiz}). 
Thus, the existence of the electrode will also significantly affect the value of the quasiparticle band gap. Note that this quasiparticle bandgap equals the difference the electron affinity and ionization energy. It is different from the optical band gap. For optical band gap one has to include the electron-hole Coulomb interaction, which partially cancells out the polarization potential effects, especially for isolated spherical quantum dots. 

In order to quantify the electrode induced localization of the electron and hole wavefunctions we define a localization function
\begin{equation}
L=\sqrt{\int \vert \psi_i(x,y,z) \vert^2 (z-z_0)^2 dr}
\label{eq: loclaiz}
\end{equation}
where the wavefunction center of mass is $z_0=\int \vert \psi_i(x,y,z) \vert^2 z dr$. It shows how spread-out the wavefunction is throughout the nanorod. This localization is presented in Figure \ref{fig: Gap_localiz} (b) and (c), for CBM and VBM respectively as a function of the electrode/nanorod overlap.  It is interesting to note that localization of the wavefunctions is different from that of the free standing nanorod already when the nanorod and the electrode are 1 nm apart. 
The localization becomes significant for both CBM and VBM when the electrode is in contact with the nanorod. 
Throughout the electrode/nanorod overlap range the electron states are less localized than the hole states due to the differences in their effective masses. As the electrode covers more of the rod length, wavefunctions become less localized since the width of the $P(z)$ confining potential is increasing. Thus there is an optimal electrode-nanorod overlap for maximally localized electron/hole wavefunctions. For the 3 nm nanorod this optimal overlap appears to be around 3 nm. 

\begin{figure}[t]
\includegraphics[width=3.2in]{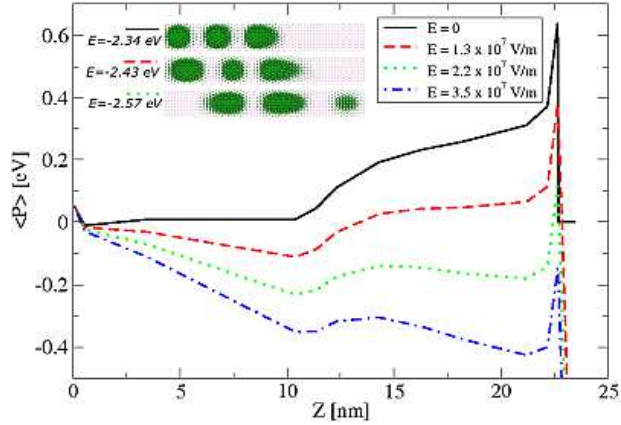}
\caption{(color online) Weighted average of the polarization function $\langle P(z) \rangle $ for 3 nm diameter CdSe nanorod, half-covered by the electrode, plotted for several values of applied external electric field. The inset shows the real space contour of the CBM 3 wavefunction along with their eigenvalues for the electric fields of 0, 1.3, and 2.2 $\times10^7$ V/m. 
The wavefunction is delocalized by the electric field, eigenvalue is shifted by 0.23 eV. 
\label{fig: E_field}}
\end{figure}
In order to make the electron mobile again, one can apply an electric field along the wire axis to overcome the wavefunction localization. 
To estimate the magnitude of this electric field 
we apply a linear potential drop to the total potential $V(r)$ in Eq. \ref{eq: quasiparticle} between the ends of the nanorod, and calculate the resulting electronic properties. This is demonstrated in Figure \ref{fig: E_field} where the weighted average of the polarization potential $\langle P(z) \rangle $ is plotted for 3 nm diameter CdSe nanorod embedded into the electrode by half of its length. The field of $2.2\times10^7$ V/m (corresponding to a bias voltage of about 0.5 V applied across the 23 nm nanorod) is needed to delocalize the CBM wavefunction. The inset shows a representative (the third) CBM wavefunction contour plots for electric fields of 0, 1.3, and 2.2 $\times10^7$ V/m (bias voltages of 0, 0.3, and 0.5 V). The wavefunction in the latter case
is dragged by the field towards the center of the rod, while its eigenvalue changes from -2.34 eV to -2.57 eV. Although it is no longer localized, it is still affected by the electrode,  the structure of the wavefunction is not the same as in a free standing rod (compare to Fig. \ref{fig: wavefunctions}a). 

In conclusion, using atomistic pseudopotential method combined with electrostatic polarization potential $P(r)$ we have demonstrated the electrode induced localization of CBM and VBM states in CdSe nanorods (representing infinitely long nanowire). The effect is surprisingly large for small (1.5 - 3 nm diameter) nanorods but becomes insignificant as nanorod size grows (6 nm and up).
The large polarization potential $P(r)$ induced by the electrode also lead to the narrowing of the quasiparticle bandgap by 0.5 eV in 3 nm CdSe nanorods. We quantify the wavefunctions localization and compute the electric fields necessary to delocalize states near the CBM. We show that the mere presence of the electrode alters the nanorod electronic properties in comparison with a free standing nanorod, and this should be taken into account in interpreting the experimental transport measurements. This is a unique nanocontact phenomenon absent in its macroscopic counterpart. 

In order to experimentally confirm the effect of electrode induced localization we propose a measurement using one of the established wavefunction mapping techniques, such as scanning tunneling microscopy (STM), along with a conductivity measurements for a series of small nanorods (1.5 - 3 nm diameter), embedded in the electrods by different degrees. 

This work was supported by U.S. Department of Energy under Contract 
No.DE-AC02-05CH11231 and used the resources of the National Energy 
Research Scientific Computing Center (NERSC).

\bibliography{electrod}
\bibliographystyle{apsrev}

\end{document}